\documentclass[draft,jgrga]{agutex}

\usepackage{lineno}
\linenumbers*[1] 

\usepackage[dvips]{graphicx}

\setkeys{Gin}{draft=false}

\authorrunninghead{Dorville et al.}

\titlerunninghead{Magnetopause: Rotational/Compressional}

\authoraddr{Corresponding author: N. Dorville, LPP, Ecole Polytechnique, CNRS, UPMC, Universit\'e Paris Sud, Palaiseau, France.(nicolas.dorville@lpp.polytechnique.fr)}

\begin{document}


%


%


\title{Rotational/ Compressional nature of the Magnetopause: application of the BV technique on a magnetopause case study}


%


%


\authors {Nicolas Dorville, {\altaffilmark{1}} 
 G\'erard Belmont, {\altaffilmark{1}}
 Laurence Rezeau,  {\altaffilmark{1}}
 Roland Grappin,  {\altaffilmark{1}}  
 Alessandro Retin\`o, {\altaffilmark{1}} 

 }

\altaffiltext{1}{LPP, Ecole Polytechnique, CNRS, UPMC, Universit\'e Paris Sud, Palaiseau, France}


%


%




\begin{abstract}

The magnetopause boundary implies two kinds of variations: a density/ temperature gradient and a magnetic field rotation. These two kinds are always observed in a close vicinity of each other, if not inseparably mixed. We present a case study from the Cluster data where the two are clearly separated and investigate the natures of both layers. We evidence that the first one is a slow shock while the second is a rotational discontinuity. The interaction between these two kinds of discontinuities is then studied with the help of 1,5-D magnetohydrodynamics simulations. The comparison with the data is quite positive and leads to think that most of the generic properties of the magnetopause may be interpreted in this sense. 

\end{abstract}


%


%



%



\begin{article}


%


%


\section{Introduction}

{\normalsize The Earth magnetopause is the outer boundary of the terrestrial magnetosphere. Outside of this boundary, the magnetosheath plasma is the shocked solar wind plasma, $i.e.$ cold and dense, with a magnetic field direction essentially determined by the solar wind one. Inside of it, the magnetospheric plasma is hot and tenuous, with a magnetic field direction essentially determined by the planetary one. For this reason, two kinds of strong gradients are observed in the same close vicinity: a compressional one (density, temperature and possibly magnetic and kinetic pressures) and a rotational one (magnetic field direction). Understanding how these two kinds of variations can co-exist is a pivotal issue for understanding the nature of the boundary. Whenever this boundary can be locally considered as stationary and one-dimensional, the Rankine-Hugoniot jump equations allow only one type of discontinuity able to ensure simultaneously the two kinds of variations: the tangential discontinuity. It implies that the normal magnetic field and flow are equal to zero in the frame of the discontinuity.
\\
There is now observational evidence (\textit{Chou et al}, 2012) that the magnetopause is not always a tangential discontinuity, $i.e.$ that the normal flow and magnetic field are not always null. There are in particular numerous observations where the magnetic field rotation corresponds to a rotational discontinuity (\textit{Sonnerup et al}, 1974). In this paper, we use the word "discontinuity'' in its usual sense in this context: a 1-D stationary layer, whatever its thickness.  
It seems that, at least when dynamical processes  takes place (due for instance to reconnection near the studied site), the compressional boundary and the rotational one can be distinguished, both propagating with different velocities with respect to the flow. Our goal is to present evidence of such a case with a detailed experimental analysis and suggest an interpretation of the observations.  In sections 2 to 4, the experimental Cluster case is presented and the two crossings are analyzed with the help of a new method presented in [\textit{Dorville et al}, submitted to JGR 2013]. In section 5, the interaction of a compressional jump and a rotational one is investigated by solving numerically the MHD equations, in order to understand to what extend the properties of both are modified by the interaction, in particular their propagation speeds. The comparison with the data and the conclusions are drawn in section 6. 
}

\section{Study of a magnetopause crossing by Cluster}
\begin{figure}[h]

\noindent\includegraphics[width=20pc]{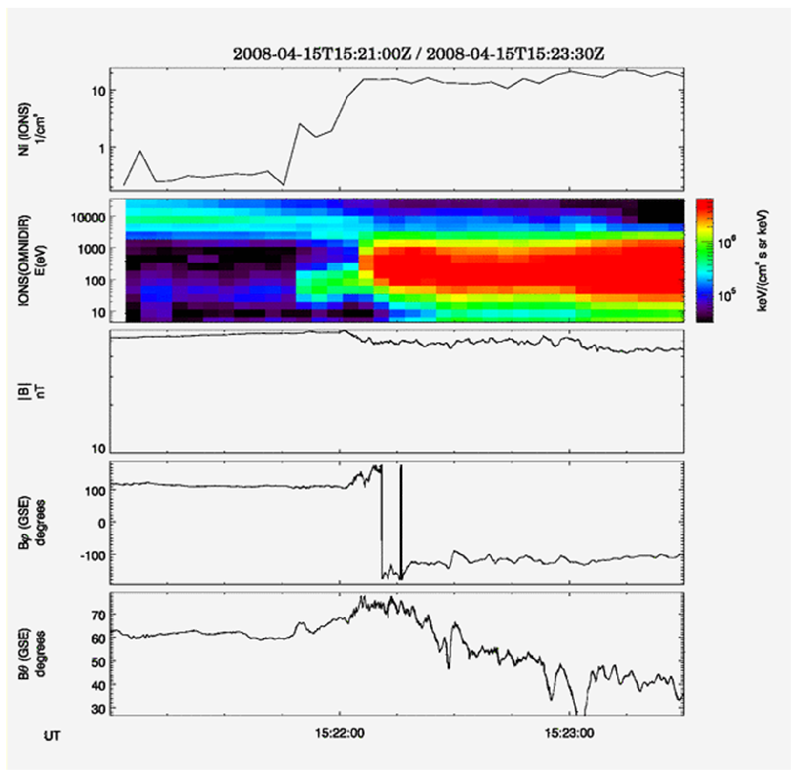}

 \caption{Density, energy spectrogram and magnetic field observed by Cluster C1 around 15:21 on 04/15/2008. The jump of density, change in plasma energy composition and rotation of magnetic field show that the satellite is crossing the magnetopause.}

 \label{CAA profiles at the crossing}
 \end{figure}

\begin{figure}[h]

\noindent\includegraphics[width=20pc]{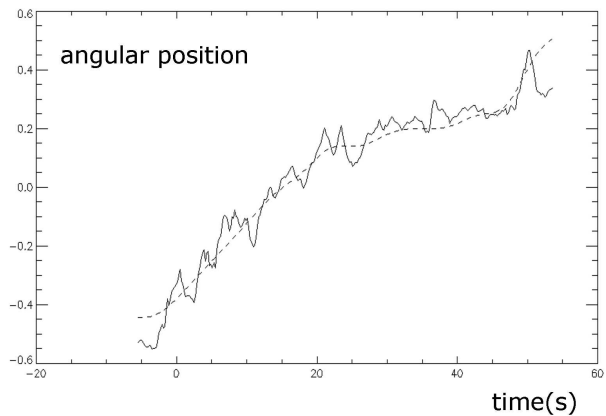}

 \caption{Observed angle on the elliptic hodogram and fit (dashed line) derived with CIS velocity by the BV method as a function of time for Cluster C1 crossing of 04/15/2008.}

 \label{angular fit}
 \end{figure}

\begin{figure}[h]

\noindent\includegraphics[width=20pc]{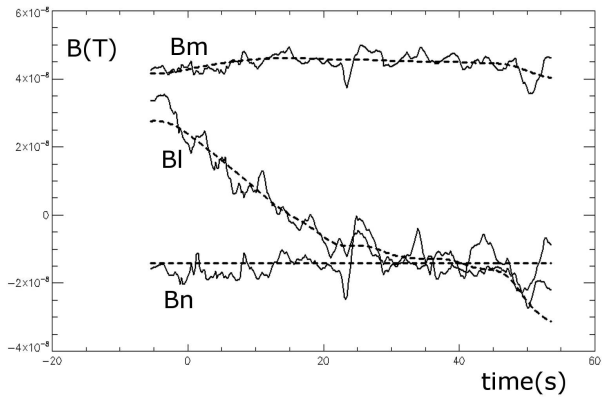}

 \caption{Three components of the magnetic field data measured by FGM and fit (dashed line) by the BV method in the magnetic field rotation region for Cluster C1 crossing of 04/15/2008.}

 \label{magnetic field fit}
 \end{figure}

\begin{figure}[h]

\noindent\includegraphics[width=20pc]{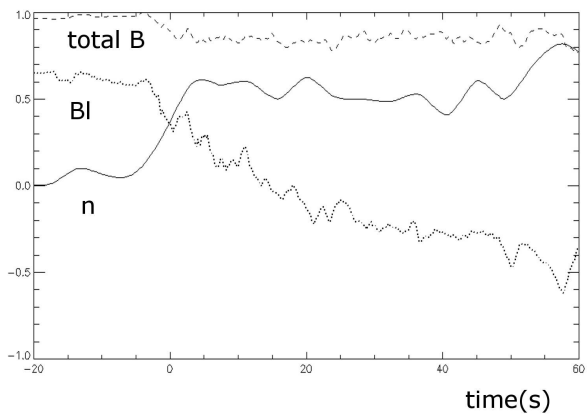}

 \caption{Time evolution of density, total magnetic field and $B_L$ component of the magnetic field during the  for Cluster C1 magnetopause crossing of 04/15/2008.}

 \label{two layers}
 \end{figure}

{\normalsize Using the single-spacecraft method of analysis described in (\textit{Dorville et al}, submitted to JGR, 2013) we are able to build a one-dimensional spatial transition parameter at the magnetopause and to analyze the small scales inside the boundary during a magnetopause crossing. This new transition parameter is expectingly close to a normal linear coordinate. The assumptions of the method is that the boundary is one-dimensional, sufficiently stationary during the crossing, and that the speed of the flow in its frame is negligible with respect to the speed of the boundary. A fit of the two temporal tangential field components with a spatial elliptic form is done, in the frame where the normal magnetic field is constant and where the normal speed of the flow is consistent with an angle on the ellipse linear with spatial position.  
\\
\\
In this part we focus on a magnetopause crossing by Cluster C1 on April 15th 2008, between 15:19:05 and 15:25:05. Fig.~\ref{CAA profiles at the crossing} shows the variation of ion density obtained by the Hot Ion Analyzer (HIA), the energy spectrum, and the variation of the Fluxgate Magnetometer (FGM) magnetic field (\textit{Balogh et al}, 1997) associated with this crossing. 
\\
For this magnetopause crossing, we are able to fit the magnetic field and find a transition parameter consistent with CIS (\textit{R\`eme et al}, 1997) ion velocity. The observed angle on the ellipse as a function of time (in black) and the angle derived by the fit with CIS velocity (in red) are plotted on Fig.~\ref{angular fit}. The fit is good on this case.  The method also provides an accurate determination of the normal direction, which is here slightly different from the MVABC [Minimum and Maximum Variance Analysis] one: ([0.98,-0.18,-0.10] against [0.96, -0.10, 0.25] in GSE coordinates). The result of the fit on magnetic field is presented on Fig.~\ref{magnetic field fit}, where time zero is the center of the large data interval and the fit is done on the time period of the magnetic field rotation. It is worth noticing that the variance on $B_m$ is not much larger than the variance on $B_n$, which makes the MVAB method inefficient. On the other hand, MVABC assumes that $B_n = 0$, which does not allow determining it either. Our method provides a value $B_n = 14nT$, corresponding to a propagation angle $\theta _{Bn} = 75$ degrees .
\\
\\
Looking inside the boundary, Fig.~\ref{two layers} presents the time evolution of density, total magnetic field and Bz. It seems clear that there are two phases in the evolution of these quantities: first, the density is rapidly increasing and the magnetic field amplitude $B$ is decreasing. Then, these quantities remains constant and the magnetic field slowly rotates on an ellipse. We will now study these two phases separately, assuming they are oriented along the same normal direction. 
}

\section{Compressional variations}

\begin{figure}[h]

\noindent\includegraphics[width=20pc]{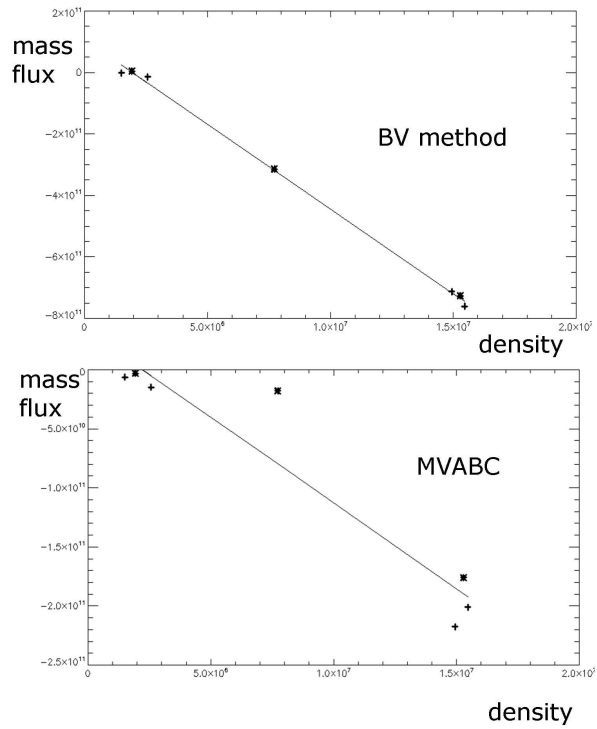}

 \caption{The normal mass flux in the shock (stars) and around it (crosses) for the BV method improved frame and in the MVABC frame. The proportionality in the BV frame shows that it is considerably better and permits to deduce the shock velocity with respect to the spacecraft.}

 \label{normal mass flux}
 \end{figure}

\begin{figure}[h]

\noindent\includegraphics[width=20pc]{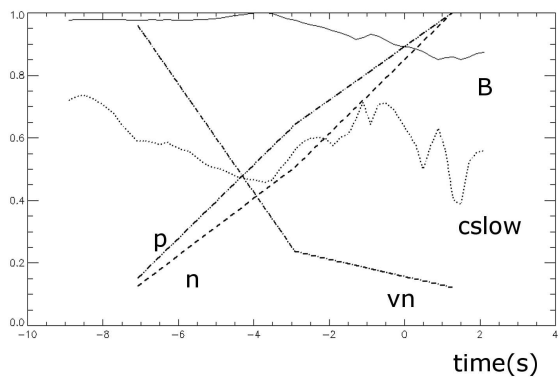}

 \caption{Normal velocity, density, pression, magnetic field modulus and slow mode velocity as a function of time for the shock.}

 \label{real data shock profiles}
 \end{figure}

{\normalsize Let us first focus on the first, compressional, phase, between t=-9 and t=2, when the density is growing and the total field decreasing. The first step to study this discontinuity is to find its normal velocity, in order to work in its proper frame. We cannot here make the simple assumption that the observed velocity is the boundary velocity because this layer is compressional and there is certainly, according with the conservation of the flux, a consistent velocity gradient inside the boundary. To determine the velocity of the boundary using the mass conservation, we plot the measured ion flux $\rho u_n$ relatively to the measured density $\rho$ Fig.~\ref{normal mass flux}, for the three CIS measurement points available inside the shock, altogether with the nearest points around the shock. This flux has to be constant in the frame on the boundary, and its variation must be proportional to the velocity of the boundary in another frame. We find that the points are aligned with a constant flux of $1.07  \pm 0.11 \ 10^{11}$ (SI) and a velocity of the boundary of $55.1 \pm 1$ Km/s, the given uncertainties  being the 1-sigma uncertainties estimated from the fits. If we restrict to the only three points in the gradient, we find that the points are perfectly aligned with a flux of $1.09  \pm 0.01 \ 10^{11}$ (SI) and a velocity of the boundary of $54.7 \pm 0.1$ km/s.
\\
\\
Here we have to notice that if we look at the same points on MVABC frame or MVAB frame, the flux is not proportional to the density. So the precision of the frame found with our method seems to be better, that is very important for such a precise treatment. Even with the small numbers of CIS points, we can be confident on this value and work on the discontinuity frame. Fig.~\ref{real data shock profiles} represents the variation of density, pressure, field modulus and normal velocity. We see that the last two quantities decrease as the others increase, that is possible only for a slow shock with respect to the Rankine-Hugoniot equations. Indeed, we find that the flow is super-slow on magnetospheric side and sub-slow on the other side, as the Cs on the curve refers to the slow mode velocity. So this discontinuity can be identified with confidence as a slow shock. }

\section{Rotational variations}

\begin{figure}[h]

\noindent\includegraphics[width=20pc]{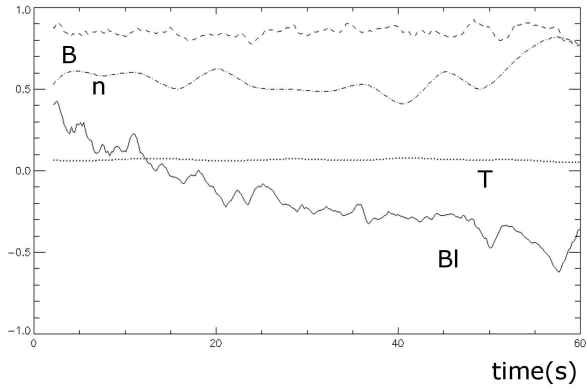}

 \caption{$B_L$ component of magnetic field, density, temperature, and magnetic field modulus as a function of time for the rotational discontinuity.}

 \label{profiles RD}
 \end{figure}

\begin{figure}[h]

\noindent\includegraphics[width=20pc]{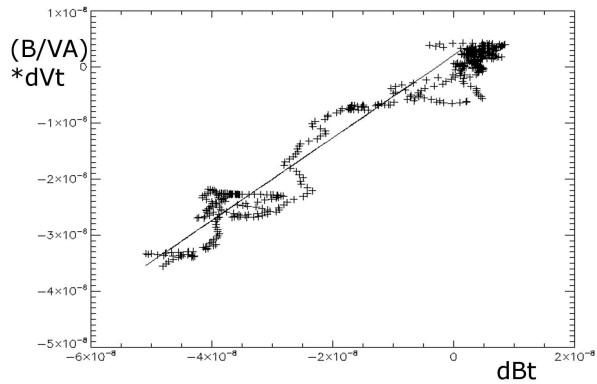}

 \caption{Variant of the Walen test for the rotational discontinuity, $\frac{B}{V_a}*dV_t$ is represented over $dB_t$.}

 \label{Walen test}
 \end{figure}

{\normalsize Fig.~\ref{profiles RD} shows the second part of this magnetopause crossing. The plotted quantities are the density, temperature, pressure and magnetic field. All seem to be constant in this part of the boundary, with only a slow rotation of the direction of magnetic field. It is also the case for the velocity modulus. These evolutions are characteristic of a rotational discontinuity.
\\
To test this discontinuity, we used a variant of the Walen test (\textit{Paschmann et al}, 2008) on Fig.~\ref{Walen test}. We plot the variations $ \delta B_t $ as a function of $ \sqrt{(\mu_0 \rho)} \ \delta u_t $ for the two tangential components. The result appears consistent with a rotational discontinuity, but with a little difference: the slope is not equal to one as it should, but to about $0.72$, $0.74$ with the correction coming from the temperature anisotropy at this place. To explain it from the composition of the plasma, we would need a mean mass of 1.8 protons. Unfortunately there is no CODIF data for C1 on April 15th, and no good resolution electron density measurement. 
\\
\\

\section{Direct numerical simulations of a slow-shock/rotational discontinuity interaction}

Let us now consider the hypothesis that this magnetopause crossing is the observation of a non stationary interaction between an Alfven wave and a slow shock. This  model should enable to describe many observations of the magnetopause, since both a compression and a rotation of $ \textbf B $ have to be present. In order to investigate this possibility, we use a one-dimensional MHD simulation code.}

\subsection{Initialisation of the shock}
\begin{figure}[h]

\noindent\includegraphics[width=20pc]{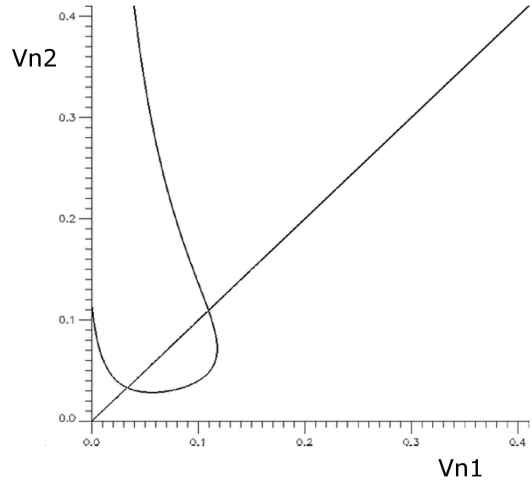}

 \caption{Possible couples of normal velocity boundary conditions allowed by Rankine-Hugoniot equations (Vn2 as a function of Vn1).}

 \label{jump conditions}
 \end{figure}

{\normalsize  We use a 1.5 D compressible MHD simulation code with periodic boundary conditions. 
The code uses a classical Fourier pseudo-spectral method to compute spatial derivatives, with an adaptative second order Adams-Bashforth time stepping for time integration. We integrate the equations for the density $\rho$, the gas pressure $P$, the velocity $u$, and the magnetic field $B$. The temperature is defined by an ideal gas law. These equations read, appropriately normalized:
\begin{eqnarray}
\partial_t \rho + \nabla \cdot (\rho u) = 0
\label{eqro} \\
\partial_t P + (u \cdot \nabla)P + \gamma P \ \nabla \cdot u = D_1 \\
\partial_t u + (u \cdot \nabla) u + \nabla (P+B^2/2)/\rho - (B \cdot \nabla) B/\rho = D_2 \\
\partial_t B + (u \cdot \nabla) B - (B \cdot \nabla) u + B  \ \nabla \cdot u = D_3
\label{eqB} \\
P = \rho T
\end{eqnarray}
$D_2$ and $D_3$ denote respectively the viscous and resistive dissipative terms, and $D_1$ the corresponding heating terms, together with conduction.
Only plane waves are considered, $i.e.$, $\nabla = (\partial_x, 0, 0)$.
The first step for this study is to obtain a sufficiently stable slow shock, in order to be able to see the interaction with an Alfven wave.
\\
To determine the needed jumps between the two sides of the shock for all parameters, we use the Rankine Hugoniot equations. We computed the different possible couples $u_{n1},u_{n2}$ for the upstream and downstream normal velocities (see Fig.~\ref{jump conditions}), for any values of the incident magnetic field and pressure.  The curve presented on Fig.~\ref{jump conditions} is for example obtained for $ P_1=0.1 $, $ B_n=0.1 $, $ \rho_1=1 $, and $ B_{t1}=1 $.
\\
We inject the chosen values $u_{n1} = 0.065$, $u_{n2} = 0.035$ in the simulation code, as the extreme values of a ramp of $u_n$ with a $\tanh kx$ profile. The width of the shock is of a few mesh points. To respect the periodic boundary conditions, we add a linear variation leading the $u_n$ function to the same value on the two boundaries. To initialize all the other quantities at all points, we directly use the conservation laws of the Rankine-Hugoniot system. 
\\
Using this initialization we obtain something which is not perfectly stationary (our equations do not take account of the code viscosity and the profiles are thus not perfectly known), but enough to study the interaction with a rotational discontinuity. Fig.~\ref{jump conditions} shows the evolution of density and fields for small times. A small perturbation of density seems to propagate alone on the right on the box (we will see that it does not change the results), and we observe a small smoothing of the gradients, but the shock is not globally moving.}

\subsection{Interaction with an Alfven wave}

\begin{figure}[h]

\noindent\includegraphics[width=20pc]{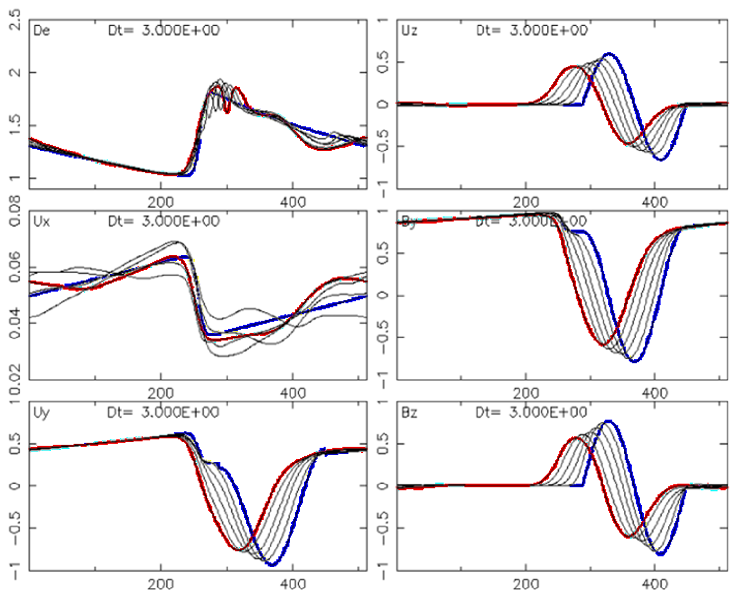}

 \caption{Early evolution of plasma parameters from the initialization (blue curve) to the last time step (red curve) when a rotational discontinuity is added in the initial condition next to the shock.}

 \label{shock+RD small times}
 \end{figure}

\begin{figure}[h]

\noindent\includegraphics[width=20pc]{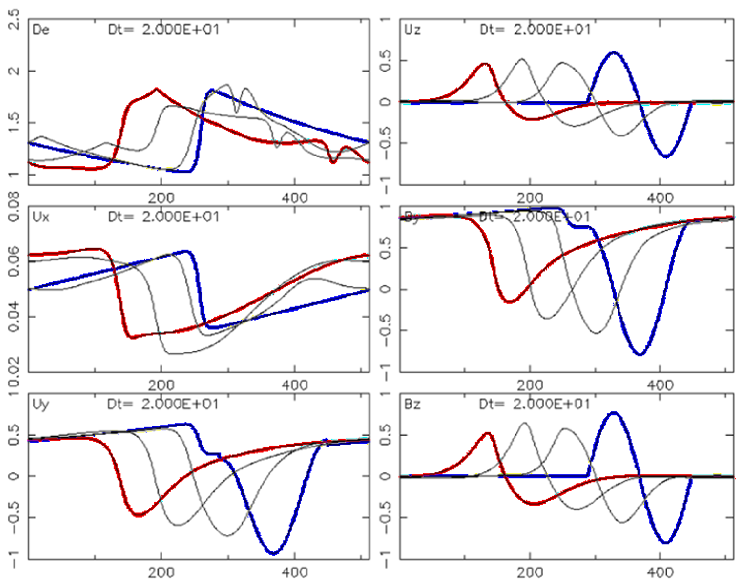}

 \caption{Long time evolution of plasma parameters from the initialization (blue curve) to the last time step (red curve) when a rotational discontinuity is added in the initial condition next to the shock.}

 \label{shock+RD long times}
 \end{figure}

\begin{figure}[h]

\noindent\includegraphics[width=20pc]{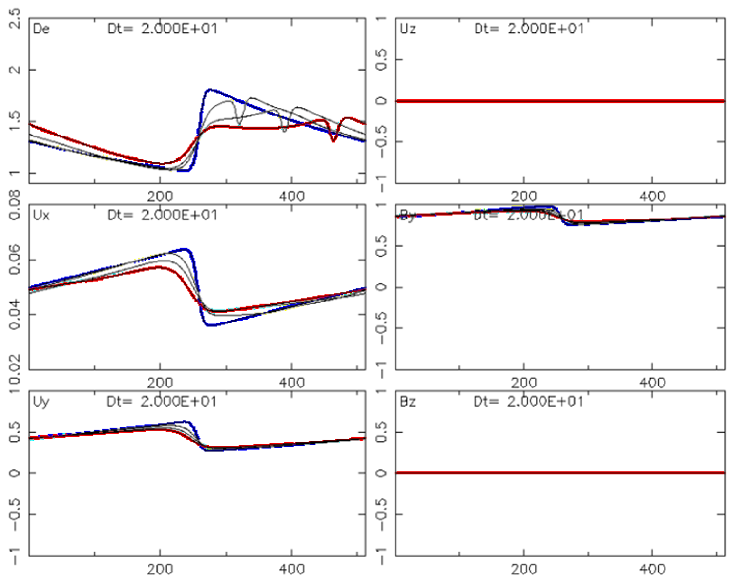}

 \caption{Long time evolution of plasma parameters from the initialization (blue curve) to the last time step (red curve) for the shock initialized alone in the box.}

 \label{shock long times}
 \end{figure}

\begin{figure}[h]

\noindent\includegraphics[width=20pc]{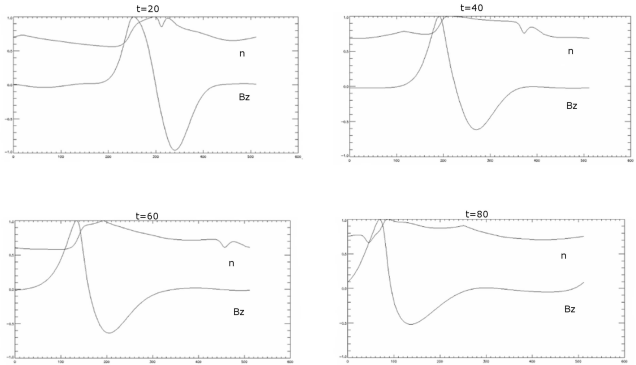}

 \caption{Shock and rotation discontinuity positions at different times. The density is used as a proxy for the shock and the Bz component as a proxy for the rotational discontinuity. The structure is not stationary.}

\label{simulation profiles at diverse times}
 \end{figure}

{\normalsize We then add an Alfven wave in the right side of the simulation box, consisting in a $2 \pi$ rotation of the magnetic field , with a Walen consistent rotation of velocity. 
The subsequent evolution after this initialization is shown on Fig.~\ref{shock+RD small times} for small times (until t=15). We see that the Alfven wave is normally propagating on the x direction from right to left and encounters the density gradient around t=9. The shock then begins to move with the Alfven wave instead of remaining steady. The imperfection on the initialization does not play any significant role in this phenomenon.
\\
This effect becomes spectacular if we look at the simulation for a longer time. Fig.~\ref{shock+RD long times} and Fig.~\ref{shock long times} show the evolution every 20 seconds during 60 seconds, respectively with the Alfven wave, and without the Alfven wave. There is no doubt that we see the formation of a slow shock/rotational discontinuity compound, the slow shock moving with the Alfven wave.
Such a structure is quite reminiscent of the so-called "double discontinuities" described in (\textit{Whang et al}, 1998). Nevertheless, these structures are usually supposed to be related to an anisotropic (\textit{Lee et al}, 2000) plasma, which is not the case in this MHD code. However, the anisotropy is said necessary to explain the stationarity of this structure. As the observed layer is not strictly stationary, even if it evolves very slowly when the two layers merge, there is no necessary contradiction. 

This kind of structure is forbidden by MHD equations if it has to be stationary, but it is not, at least at times we can reach with the simulation, as shown on Fig.~\ref{simulation profiles at diverse times} that shows the respective positions of the Bz and density gradients at different times. The shock affects the shape of the rotation of magnetic field (even if a Walen test can show it is still Alfvenic), and the compressional part position seems to be oscillating. 
\\
}
\section{Conclusion}

We have presented a case study of a magnetopause crossing by Cluster spacecraft. This example is atypical in the sense that the density/ temperature gradient is clearly separated from the magnetic field rotation layer. This allows to investigate separately the two layers and determine their nature. Using a new method, the normal direction is determined with a good accuracy and the normal components $B_n$ of the magnetic field and $u_n$ of the velocity can be measured in a reliable way: the two layers are not tangential. In addition, the method allows plotting the different profiles in function of a transition parameter which is a trustworthy proxy of a normal coordinate. 
The density/temperature gradient layer could so be identified unambiguously as a slow shock, while the rotation layer has all the properties of a rotational discontinuity, except that it may propagate at 0.7 $V_A$ instead of $V_A$. A 1,5 D MHD simulation has been performed to investigate the interaction between a slow shock and a rotational discontinuity when they meet. The results are quite comparable with the data : The shock stays attached to the rotational discontinuity, at least before viscosity kills all the gradients in the simulation, and keeps shock properties as the second part of the discontinuity is still rotational. The structure is not stationary and so stays compatible with MHD equations. We can conjecture that this kind of interaction between compressional and rotational features is a generic feature of the magnetopause.

\begin{acknowledgments}
The authors would like to thank the CAA and all Cluster instruments teams for their work on Cluster data.
\end{acknowledgments}



%


\end {article}


 \begin{figure}[h]

\noindent\includegraphics[width=20pc]{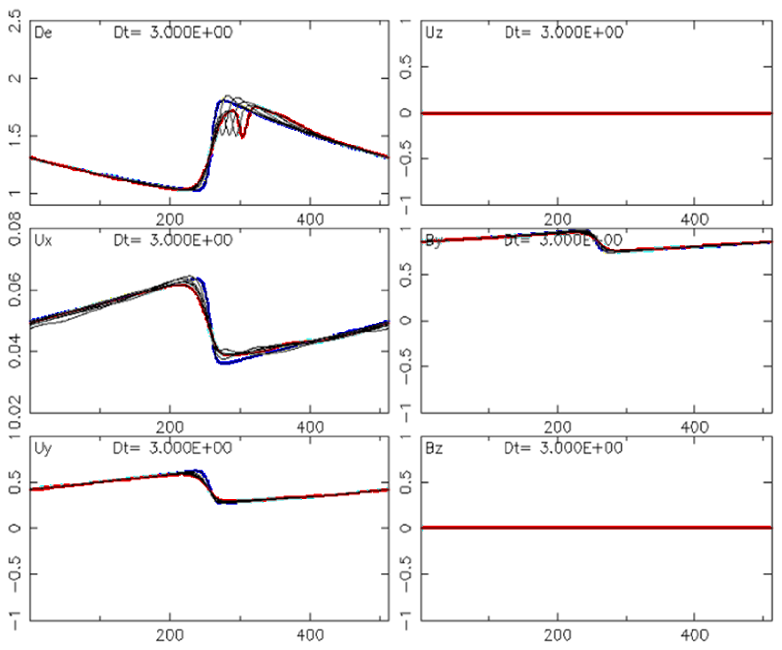}

 \caption{Early evolution of plasma parameters from the initialization (blue curve) to the last time step (red curve) for the shock initialized alone in the box.}

 \label{shock small times}
 \end{figure}


\end{document}